# For a Semantic Web based Peer-reviewing and Publication of Research Results

Philippe A. MARTIN

EA2525 LIM, University of La Réunion (+ adjunct researcher of the School of ICT, Griffith Uni, Australia)
Sainte Clotilde, La Réunion, France

## ABSTRACT

This article shows why the diffusion and peer-reviewing of research results would be more efficient, precise and relevant if all or at least some parts of the descriptions and peer-reviews of research results took the form of a fine-grained semantic network, within articles or knowledge bases, as part of the Semantic Web. This article also shows some ways this can be done and hence how research journal/proceeding publishers could allow this. So far, the World Wide Web Consortium (W3C) has not proposed simple notations and cooperation protocols – similar to those illustrated or referred to in this article – but it now seems likely that Wikipedia/Wikidata, Google or the W3C will propose them sooner or later. Then, research journal/proceeding publishers and researchers may or may not quickly use this approach.

## 1. INTRODUCTION

Scientific knowledge diffusion and validation currently essentially relies on the writing of mostly informal research *articles* and mostly informal peer-reviews of them. In this article, "formal" means "with a unique meaning and, generally, some logic-based structure or partial/complete definitions that a software can exploit". According to [1], researchers "know that the [current] system of peer review is biased, unjust, unaccountable, incomplete, easily fixed, often insulting, usually ignorant, occasionally foolish, and frequently wrong". [2] lists many studies showing how broken *article peer reviewing* can sometimes be. Indeed, judging a whole informal article (its significance, presentation, …) is more difficult and more background/personality/goal dependent – thus, in a sense, more arbitrary – than correcting and giving arguments for or against the veracity and significance of each single sentence or idea about a research work. This article makes the case that both peer reviewing and knowledge diffusion would be more efficient, precise and relevant if all or at least some parts of the research results and their peer-reviews took the form of a fine-grained semantic network, within articles or knowledge bases, as part of the Semantic Web (SW) [3]. The more fine-grained, the better, but this would be up to each author and reviewer. This would help retrieve, compare and relate research results and hence would also help reduce the large number of redundancies between the research results within the huge amount of research articles (about 1.3 million research papers each year according to a study made in 2009 [4]), most of which being seldom read of cited (according to [5], 90% of journal articles are never cited). Furthermore, the work of a reviewer would then be close or identical to that of an author, and hence could be similarly exploited and rewarded.

Given the well-known limits of lexical-based approaches for information retrieval, many researchers have used logic-based languages and semantic relations (e.g., specialization/part-of/spatial/temporal relations) for organizing and indexing or partially representing some meanings of some information. The more precise and homogeneous the representations, the better "knowledge retrieval, comparison, inferencing and validation" can be. Since 1998, the World Wide Web Consortium (W3C) has standardized some formal languages and "formal vocabularies" ("ontologies") to help the development of the SW (the formal part of the Web and the part indexed by this formal part). Many research organizations use these languages to publish their databases and allow them to be queried and exploited via languages more powerful than SQL. This therefore also eases the integration or cross-querying of databases. However, as shown in Section 2, the W3C has not yet proposed a formal *notation* sufficiently expressive and high-level to enable researchers to i) represent and semantically organize the kinds of natural language sentences, ideas or "know-ledge" that can be found in research articles, nor ii) annotate published "data" (the elements of the published research databases) with such "knowledge". The SW community has only created an ontology of some SW related research domains [6]. Few research works are aiming at creating such notations, mainly [7], [8] and [9] (other works on "controlled languages" led to notations which, for many purposes, are not formal or expressive enough). As illustrated below, FL [9] is the simplest textual notation for simple knowledge representation cases, the most visually structured, and the most flexible: it draws from various families of notations, allows to mix them, and its syntax will soon be dynamically adaptable by its users. Hence, FL is used in Section 2 to illustrate how some content of an article can be formally or semi-formally represented in a rather easy way as well as in an incremental and cooperative way.

The W3C has also not proposed *cooperation protocols* nor a general *top-level ontology* to ease *knowledge sharing*. However, others have made such propositions. Section 3 lists requirements and solutions for researchers (authors and reviewers) to represent and relate their knowledge (ideas, arguments and objections for them, source facts/data, techniques, tool features, etc.) in an organized way into one knowledge base (KB) or several inter-related KBs managed by research journal/proceeding publishers, research communities or other organizations. Section 4 summarizes a general framework for evaluating the knowledge and knowledge authors of such KBs. The recent but now official interest of Wikipedia/Wikidata and Google to use SW techniques will probably lead them to adopt – and thus, popularize – similar protocols, general ontologies, notations and perhaps, evaluation frameworks.

Section 5 elaborates on the advantages and drawbacks for authors, reviewers and publishers to adopt KB based techniques – within KBs and/or articles – as a complement to traditional techniques for research knowledge peer-reviewing and diffusion.

## 2. REPRESENTING KNOWLEDGE

KB-based knowledge sharing and reviewing is about *representing relations* between elements (sentences, terms for concepts and relations, …) and hence also defining them. The more defined these elements and relations, the smaller the elements, and the greater number of relations, the better. There are many kinds of knowledge representation languages (KRLs), with different logic models (and hence expressiveness) and kinds of notations. Graphic notations are nice to look at but are not concise, are often poorly expressive, have no standard format and using them is time-consuming. Thus, in the same way that graphic-based programming languages are difficult to use for developing big applications, graphic-based KRLs are difficult to use for developing large or complex KBs. Most textual notations for KRLs are either relation/predicate-based, English-looking, frame/graph-based, XML-based or HTML-tag-attribute-based. The de-facto standard for the first kind is KIF [10]. The W3C proposes a standard for each of the last three kinds; respectively: SPARQL (+OWL) Update/Query, RDF(+OWL)/XML (in this term, "/" means "linearized with") and RDFa. FL covers the first three kinds. Below are representations of the sentence "there is a man named 'Joe' that has at least 1 leg"; in addition to predefined terms, these representations use only terms defined by a person identified as "p" (this is a shortcut; a whole URL can also be used):
– in FL:  `a `p#man with p#name "Joe" ´ has for p#part a p#leg´
– also in FL:  `a p#man  p#name: "Joe",  pm#part: a pm#leg´
– in KIF:   (exists ((?m p%man) (?l p%leg))
            (and (p%name ?m "Joe") (p%part ?m ?l)) )
– in SPARQL: insert {?m  a p:man; p:name "Joe"; p:part [a p:leg] }
– in RDF/XML: <p:man p:name="Joe"><p:part><p:leg/></p:part>
              </p:man>
– in RDFa:  <div typeof="p:man">
              <span property="p:name">Joe</span>
              <span property="p:part"><span typeof="p:leg"/>
              </span> </div>

Out of these five KRLs, only FL and KIF have all the **low-level constructs** necessary for representing common natural language (NL) sentences (and, more generally, not just "simple kinds of knowledge"): a second-order logic syntax, a first-order-logic-at-least model, meta-statements (i.e., the possibility to write statements about statements), contexts (meta-statements that give conditions without which the inner statements may be false, e.g., spatial/temporal/modal conditions), the possibility to *define* kinds of quantifiers (e.g., numerical quantifiers) or collections (sets, alternatives, distributive sets, …), and distinct constructs for "defining" and "universally quantifying" (making this distinction is useful when the KRL is not based on a second-order logic model). SPARQL has a cumbersome syntax for meta-statements (and only for simple kinds of them), and such simple meta-statements will soon be re-introduced in RDF(+OWL)/XML. None of the other low-level building blocks are yet provided by the W3C notations. E.g., the sentence "Dr. Foo *believes* that *in* 2012, France, *at least 78%* of healthy birds *were able to* fly" can be represented in FL and in KIF but, because of any of its words in italics, it cannot be represented in the other above cited KRLs. In FL, still using only terms from "p": `p#DrFoo p#believer of ` ` `at least 78% of p#healthy p#bird can be p#agent of a p#flight´ with p#place p#France´ with p#time 2012´ ´.

Out of the five above cited KRLs, only FL has **high-level constructs** necessary for people to represent common NL sentences in easier and more normalized ways, hence in more correct and automatically comparable ways. For example, constructs for numerical quantification (e.g., `at least 78% of´ and `between 2 and 4´), valuation (e.g., `a p#cat with p#weight 1.45 p#kg´ can, if needed, be translated into longer forms such as `a p#cat with p#attribute a p#weight that has for p#measure a p#Measure that has for p#unit a p#kg and for p#value 1.45´), attribution (e.g., `a p#red p#car´ can be translated into `a p#car with p#color a p#red´ or `a p#car with p#attribute a p#color that has for p#measure a p#red´), etc. The RDF+OWL model – used by the W3C KRLs – not only lacks essential low-level constructs (as above summarized) but has very cumbersome constructs for very common notions. E.g., the sentence "(it happens that) men have at most two legs" – which in FL can be simply represented by `every p#man has for p#part at most 2 p#leg´ – has to be represented as follows in SPARQL:
  insert { p:legAsPart rdfs:subPropertyOf p:part;  rdfs:range p:leg;
           p:man  rdfs:subClassOf  [a owl:Restriction;
           owl:onProperty p:legAsPart  owl:maxCardinality 2] }
Actually, strictly speaking, this last representation means that "by definition of p:man, any p:man must have at most 2 legs", which is not what the original sentence really meant. In RDF/XML and RDFa the representations of this sentence are even more difficult to write and read. To sum up, most sentences found in research articles – and many/most of the ideas/knowledge they describe – cannot be written with the RDF+OWL model, or an XML/HTML-based syntax, or the current syntax of SPARQL even if more powerful models are used. Even if they can be written with such syntaxes and model, displaying them as such (without translating them into higher-level notations) often makes them difficult to understand. The use of an high-level syntax which, like FL, can represent a lot of knowledge in a concise, uniform and visually structured way, is needed for people to explore and understand relatively complex KBs in reasonably efficient ways and update them in relevant ways for knowledge sharing purposes. Unfortunately, most knowledge editors – or, more generally, knowledge based tools – do not yet use such high-level notations and hence only show relations directly connected to a particular object. Using such knowledge editors to update a large or complex KB is somewhat analog to developing a large program with a line editor instead of a full "text editor". The notion of "being concise and visually structured" is illustrated by the next two FL examples. In these examples, when not specified, the creators/sources of the terms and relations is assumed to be "p" and hence is left implicit. It should be reminded that this section is not "trying to sell FL", it is only using it to give an idea of the kind of work and notations that KB-based knowledge sharing and review requires.

The next example shows various kinds of relations from a term. It should be read: i) *any* instance of the type information_sharing (i.e., any process of this type) *has for* subtask *0 to many* instances of information_diffusion, *0 to many* instances of information_retrieval and *0 to many* instances of information_validation, a type which has for subtype peer_reviewing, and ii) *any* instance of the type information_sharing has for object (i.e., has for input and/or output) *1 to many* instances of information_object, has for (related) rule something which informally can be expressed as "the more precise … and reuse",  and may have other relations to other objects. The comment (after "//") is for presentation purposes only.

```
information_sharing
  subtask: information_diffusion  information_retrieval
           (information_validation  subtype: peer_reviewing),
  object: 1..* information_object,
  rule: "the more precise the shared  information_object, the better for
         information sharing and re-use";  //argumentation structure below
```

The following example shows how sentences can be inter-related or annotated. The first and the third sentences (in this example) are semi-formal, i.e., have both formal and informal terms. The third has a formal structure (only one relation name is informal): it uses back-quotes and right-quotes for sentence embedding. This example shows how people can progressively and collaboratively formalize, annotate and refine knowledge, correct it without removing it (thanks to relations such as p#corrective_precision) and argue or object it. This example shows that, in FL, one way to associate meta-information to a relation (and hence to the sentence constituted by this relation) is to put these meta-information in the __[...] construct after the destination of the relation. Here, this illustrates the use of an objection relation *on* a relation (not on its destination) to represent an objection on the *relevance* of the use of a sentence as an argument/objection (not on the veracity of this destination sentence). Few argumentation systems allow to make such a distinction and thus, few do not lead to biased information. ArguMed is one of the exceptions. ScholOnto [11], despite being ontology-based and intended for organizing scholarly claims, is not an exception. These systems are hypertext ones: they do not permit people to represent knowledge in formal ways. Like most current KB systems, they do not advocate (nor control the following of) knowledge design best practices [14] that lead to more precise and normalized knowledge, thus avoiding redundancies and easing its understanding and exploitation. E.g., one rule for avoiding argumentation structures to turn into "spaghetti-like networks" when they grow is, whenever possible, not to use p#objection or p#correction relations but p#restrictive_correction or p#corrective_generalization relations and then, as meta-information on them, use p#argument relations. This also improves the precision and acceptance of the corrections.

```
"the more precise the shared pm#information_object, the better for
    information sharing and re-use"
  argument: ("the more precise an information object, the easier it is
              to handle automatically and correctly"
       specialization_or_equivalent_object:
          p#` if `?o1 specialization: ?o2´
              then `?o1 "is easier to handle correctly than":
                    ?o2´ ´),
  objection: (p#"the more precise an information object ,
              the more difficult this object is to write"
       corrective_precision:  //by "p" (the default source here)
              "giving more precision takes more time to
               write and is sometimes more difficult"
  ) __[ author: oc,   //but the author of the next objection is "p":
        objection: "someone spending time to share information
                    generally does not mind spending a bit more
                    time to make it more accessible and used" ];
```

### 3. SHARING KNOWLEDGE

**Knowledge sharing and organization within a KB**

For people – researchers, lecturers, engineers, … – to be willing and able to store and relate their knowledge in a precise, organized and scalable way into the shared KB of an organization, community or publisher, the KB management system (KBMS) must have *at least* the next listed features [9] [12-15].
1) The KBMS must offer expressive and high-level notations for users to add or query knowledge, and define filters to see only what they wish when browsing or querying the KB, e.g., only knowledge specializing a given formal sentence and created by certain kinds of persons (e.g., persons having authored something that some user finds highly original). At least one notation should allow the presentation of knowledge as a unique graph (so far, it seems that only FL allows this). Query results should at least show the specialization relations between the results (and hence they should be organized into a specialization hierarchy) and, from them, exploration to related objects in the KB should be possible.
2) The KBMS should propose a specialization relation that allows formal and informal knowledge (terms, sentences, …) to be organized into a single specialization hierarchy and thus i) to be managed via similar semantic procedures, and ii) to have a unique place in this hierarchy. This ensures that every piece of knowledge can be compared with every other one, at least according to specialization relations. Thus, this helps reduce implicit redundancies and is a requirement for scalability [16]. [9] proposes such a relation and associated procedures.
3) The KBMS must have a *KB editing protocol* which does not accept knowledge addition or removal when this violates some rules imposed by the KB owner or agreed to by the user, e.g., the following of some knowledge design best practices (some that the provided high-level notations cannot enforce or encourage) [14]. An important rule is not to remove or modify someone else's knowledge. Another one is not to introduce an implicit inconsistency or redundancy in the KB. This does not prevent a user to disagree with another one but the kind of disagreement (and, if needed, why they disagree) must be stated explicitly, at least by using relations such as p#correction, p#corrective_generalization or p#corrective_precision, as previously illustrated. Using these last two kinds of relation is better since they participate to organizing knowledge into the general specialization hierarchy. Thus, the users do not have to agree on terminology and beliefs but still have to relate their knowledge. [9] and [13] propose a much more detailed set of rules to keep the KB organized and free of *implicit* inconsistencies or redundancies, once they are detected by the KBMS or by users. In case of removal, some these rules involve an automatic "cloning" of the deleted knowledge, i.e., the attribution of its ownership to another user who relied on it. Freebase, the KB which Google exploits and allows people to contribute to, also uses a "loss-less approach" for knowledge sharing but cannot enforce the use of corrective relations since most of the content of Freebase is automatically extracted. The above described approach can work with both formal and informal knowledge, and hence could be applied to classic wikis and semantic wikis. Along with the previous points, it avoids the need for the KB owners to impose arbitrary restrictions on the content of the KB and then constantly enforce them for each new piece of knowledge. Thus, this avoids one bottleneck of classic KB sharing. Finally, this approach can be combined with other approaches for knowledge sharing, and the corrective relations can be exploited for knowledge selection, e.g., one may choose to see, believe or re-use only the knowledge that has *not* been "corrected" *and* that are from certain kinds of users.
4) To guide users and alleviate their workload, the KBMS must provide a general ontology of natural language organized by general top-level ontologies. [12] presents the core of one such ontology, formed by loss-less integration and completion of other ones. It helps the detection of inconsistencies but is not a big help in the avoidance of implicit redundancies. Fortunately, there are more and more interconnections between the major large general ontologies. This will help forming (a) better one(s).
5) To guide users, the KBMS must also provide a top-level ontology for its domain. [15] presents the core of a top-level ontology in Knowledge Engineering.

**Knowledge sharing between individual/community KBs**

Ideally, it should not matter which KB a person (researcher, ...) chooses to query or update first: object additions/updates made in one KB should be replicated into all other KBs that have a scope which covers these objects. Idem for queries when this is relevant. In other words, ideally, the (Web-accessible) KBMSs of different organizations or persons should be able to interact for their KBs to be "views" on one global virtual KB. The approaches used by current distributed systems (including knowledge-oriented P2P ones) are not shared-KB-based enough to be extended for implementing the above vision. However, KBMSs can still achieve it if, for every term T a KBMS stores, it either

1) has a Web-accessible formal description specifying that it commits to be a "nexus" for T, i.e., that i) it accepts – and try to gather – any statement S on T, or ii) it associates to T the URLs of KBMSs permitting to find or store any statement on T, or

2) is not a "nexus" for T, and hence it associates to T either i) the URLs of all KBMSs that have advertised themselves as nexus for T, or ii) the URL of at least one KBMS that stores these URLs of nexus KBMSs for T.

Thus, via forwards between KBMSs, all knowledge using T can be added or found in each nexus for T [9].

**Comparison with other approaches**

The current SW is mostly composed of small, single-authored, (semi-)independently authored, heterogeneous, poorly organized and poorly inter-related static RDF/XML files [17]. Most current SW related tools focuses on helping align, merge – and, more generally, exploit – such KBs in order to perform or ease automatic reasoning. Thus, they are intended for applications, not knowledge sharing, and they alleviate the difficulties caused by the lack of inter-relations between KBs. However, the outputs of these tools are often new "static KBs poorly inter-related with their source KBs" [17], i.e., these outputs are not inserted into shared KBs. Thus, these tools also contribute to the above cited difficulties. Using the outputs of these tools as inputs for the above described approaches is difficult – and has to be mostly manual – since much of the required information has not been made explicit by the creators of the source KBs. Some other current SW related tools are "personal KB editors" or "shared KB servers/editors" but they also create "new KBs" and do not yet use KB editing protocols nor inter-KB knowledge replication protocols.

## 4. EVALUATING KNOWLEDGE AND AUTHORS

Collaboratively finding arguments and objections about the originality, veracity, current or future significance, …, of one simple statement is not easy, is sometimes background/personality/goal dependent but is often fruitful. The resulting argumentation structure is an organized multi-viewpoint state of the art on one object (idea, method, …), something which is difficult to come up with, even by reading many articles. Thanks to it, each person can make his own judgment based on his goals and their associated constraints. Rating one simple statement (with qualitative/quantitative values) according to any of the above cited criteria is somewhat arbitrary. Rating a group of (formal or informal) statements is even more arbitrary. Combining the various rates (for the various criteria, given by one or several persons) is again even more arbitrary, and semantically meaningless unless one group of statements is better than another one for every criteria. It is also fruitless except for selection purposes.

With the proposed KB-based knowledge sharing and reviewing approach, arguing or rating groups of statements becomes unnecessary.

However, many persons will still want statements to be rated, and these ratings combined, e.g., for ordering statements or their authors according to criteria (originality, …) and combinations of them. This may for example be useful for display purposes or grant attribution purposes, even if the combinations are clearly "semantically" meaningless (they are not "mathematically" meaningless). Hence, the rating of one statement according to one criteria by many persons should, as much as possible, be automatic and knowledge-based (typically, it should be based on how this statement has been rated and argued for and against for this criteria by each of the persons, e.g., by a recursive exploration and weighting of each argument and objection in its argumentation structure). Even more importantly, the procedures for rating one statement for one criteria, and then for combining all the ratings, should not be predefined in a KBMS: each user should be given the possibility to define or parameterize parts or all of the procedures. [9] proposes a "default measure" for the "average usefulness" of an object (term or statement) based on a recursive exploration and weighting of the users' individual evaluations of this object and, to a small extent, the "average usefulness" of these users. This last one is similarly derived from the way their objects have been evaluated and from their participation to evaluate other users' objects (as an incentive to do such a work). These "usefulness" measures are completely – and necessarily – arbitrary and semantically meaningless (a square root function is even used at one stage). However, they are meant to be parameterized by each user if he wishes to. More importantly, since they can be used to display statements with bigger or smaller fonts, they are a default way for users not to be bothered by statements with low "average usefulness" and this should be an incentive for authors to create statements of better "usefulness" (as defined by the default measure). This approach may be seen as offering the beginning of a technical grounding for the very general "model of discursive practice" of Brandom [18]. It is also a flexible and scalable alternative to the Co4 protocol [19] in there is no shared KB (each user has a personal KB) and the protocol derives a hierarchy of more and more consensual KBs (hence, an ordering of statements based on how consensual they are) based on exchanges between users and similarities between their KBs.

## 5. SUMMARY OF ADVANTAGES AND DRAWBACKS

The more sentences or relations a chunk of information includes, and the less formal it is, the more difficult this chunk is to relate to another chunk of information via a precise semantic relation. In other words, the more difficult it is to state things precisely and correctly about such chunks, to compare, index or organize them, and hence to retrieve them. *From this viewpoint,* for the purposes of sharing, validating or arguing for/against some pieces of knowledge, the *"writing, peer-reviewing and publishing of informal research articles" is the approach that inherently is the least efficient* and which most leads to redundancies as well as imprecise, incorrect or arbitrary statements ("arbitrary" in the sense of "dependent of each person's goals, preferences and cultural background"). This approach makes peer-reviewing difficult. E.g., how to detect plagiarism and then (except for clear cases) how to judge in a non-arbitrary way what is (self-)plagiarism (versus, original) or not? This approach also makes knowledge retrieval, understanding and learning difficult since it involves reading, cross-checking and remembering the content of many documents. It also makes knowledge writing difficult since it often leads to "space constraints" and since it makes "presentation" important: which elements to introduce and to which point, in which order, with which informal structure, etc. *It makes knowledge writing an art.*

*When representing knowledge in a "shared" KB* (e.g., a KB of a research community, interconnected KBs, or even the whole Semantic Web) *for general knowledge sharing purposes* (as opposed to application development purposes), *one* can and should re-use or refer to as much existing knowledge as he can but he *only has to think about correctness, precision and sharability*. This means that one only has to think about representing as many correct relations as he can (those that the KB system detects as already directly or indirectly represented should be rejected by this system). Best practices, protocols and "incentives for knowledge inter-linking/re-use" will – when more developed and popularized – help and lead each KB contributor to think about correctness, precision and sharability, i.e., to make semantic relations explicit within his knowledge as well as from/to other people's knowledge, e.g., via explicit argumentation structures. The bigger the shared KB, the more the user will be led to remove or correct and explain potential redundancies or contradictions and, more generally, led to provide information that only he can easily provide. Thus, a contributor to a shared KB can – and will be encouraged to – provide more information than in a research article, e.g., *more technical* information. Indeed, at least in non-theoretical Information Technology related research, based on the kinds of arguments that one can find in reviews of research articles, one may think that to get one's article published it may be safer not to describe or use things that require some focus (i.e., more than a cursory reading) to be understood or at least not mis-interpreted. Examples of such things: i) non-mainstream approaches, ideas or formalisms (e.g., KIF, now that mainstream formalisms are much less expressive), ii) arguments against mainstream approaches, ideas or formalisms, and iii) technical information.

Furthermore, a *KB user does not have to write a whole new* (and relatively self-contained) *article for each new advance in his research* (doing so creates redundancies) *and does not have to think about presentation (element order, space restriction, …) or the cultural background of its potential readers*. Indeed, knowledge representations can be selected and displayed with great flexibility, even automatically according to the knowledge and preferences of each user if some of them have been represented too. Furthermore, navigating along relations of a well-organized semantic network permits someone to quickly find and compare what he wants.

Finally, with the proposed KB-based approach, there is no more a difference between an author and a reviewer. Both can be rewarded, and evaluation schemas can take advantage of the "fine-grained nature, precision and semantic inter-relation" of the various contributions, as illustrated in the previous sections.

So far, for the publication and peer-reviewing of general research results, there was no alternative to the classic informal document based approach. This will still be the case until Semantic Web approaches and techniques have been made more popular – at least in some researcher communities – by organizations such as Wikipedia, Google or biology-related organizations. Then, will KBs be quickly adopted for the publication, peer-reviewing and organization of general research results (i.e., ideas and theories, not data)? Some advantages of doing so have been listed by the previous paragraphs and they seem important for coping with the increasing number of researchers and research outputs. Strong obstacles are that i) most researchers would have to learn at least some "basic notions and formal terms" for knowledge representation, and ii) they are often not inclined to learn them, nor to express themselves in such a semantically structured way. Many researchers will also not be interested in doing so because they publish articles easily, or they regularly manage to publish articles which have many redundancies with their previous articles or other persons' articles, or they regularly manage to publish articles whose content will become much more easily recognized as hollow, inconsistent or "incorrectly argued for" when represented and organized using semantic relations. It is also true that some kinds of knowledge are difficult to represent or organize semantically, even in a semi-formal way, but this is generally because doing so enforces the representation of conceptual distinctions which bring delicate or problematic questions to the forefront. Another big obstacle may be the slow official recognition of ways to evaluate researchers based on their contributions to KBs. Many researchers in biology related domains already face a similar problem regarding their contributions to databases. One incentive for researchers to add to databases or KBs instead or in addition to research articles is that this makes their data or knowledge more easily accessible and, in the future, more correctly and precisely reviewed than if it was presented informally in a research article. One incentive for researchers to make such reviews will be to give less arbitrary reviews and to be rewarded as authors for them. One incentive for publishers to enable the publication of research results via KBs will be i) to have researchers and engineers paying to access well-organized KBs, or ii) to be able to associate relevant advertisements to elements of the KB (since these elements will be precise). On the other hand, publishers may then less be able to sell journals or proceedings.

If KBs can be used for peer-reviewing and publishing research knowledge, publishers or other organizations are hopefully likely to also allow research articles to include formal representations (in a particularly readable format then) as well as references to pieces of knowledge in these KBs or in databases. At least in mathematics, this is often already the case. In other domains – including Information Technology – this would ease certain aspects of the writing of articles since i) these articles would not have to introduce notions that are introduced in a KB, and ii) within the formal parts, "presentation" (element order, …) and "taking into account the reader's background" would be less of a concern. Furthermore, researchers would be able to easily publish these parts in KBs too and, at least for the content of these formal parts, would have less chances to be reviewed in arbitrary ways or then would have more chances to have a semi-formal discussion on precise points with the reviewer. This would likely be advantageous for both the author and the reviewer. An incentive for publishers to permit such formal parts and references in journal articles is that this may attract more research submissions. publishers may also try to only allow references to knowledge in the publisher's KB: this may or may not be an incentive for authors to contribute to this KB and for the readers of these "articles with formal parts" to access this KB. The interest of researchers to buy such "articles with formal parts and references to KBs" is also in question: they may be interested "in not having to always read introductions to techniques they know" and by the precision and organization of information and arguments, or they may dislike formal formats and having to make occasional searches in KBs for getting further information. In other words, do the informal, linear and self-contained natures of many current research articles inherently suit many persons or are they an heritage of the time when research articles could only be "research papers"?

## 6. CONCLUSIONS

The current use of research articles – and peer-reviews of them – as the principal way to diffuse and validate research result is known to have many problems. This article identified the main cause of these problems as being the use and review of *whole* informal documents. However, only variations of the "research result diffusion and peer-reviewing" processes seem to have been discussed or tried, e.g., the "open access to articles" (after some period of time, or directly if article authors pay for it), "shared reviewing of articles" (amongst a consortia of journals), "public reviewing" (anyone can submit a review), "non-anonymous reviewing" and "open access to reviews". [2] proposes combinations of such variations in order to reduce their respective disadvantages. The improvements brought by these variations are interesting but, from the viewpoint developed in this article, can only be limited. E.g., enforcing, proposing or lifting anonymity for authors and/or reviewers may or may not have some advantages (depending on who is reviewed and who does/would review) but cannot really solve the above cited problems since they are caused by the fact that whole informal articles, not individual semi-formal sentences, are reviewed and published. In the approach advocated by this article, anonymity is fortunately rarely useful anymore.

To improve those processes, this article advocated a "cooperatively built KB" based approach. The sections 2, 3 and 4 illustrated the requirements and solutions for enabling and encouraging, respectively, "precise and/or correct knowledge representation", "knowledge sharing" and "knowledge/author evaluation". However, applying such techniques to "diffuse and validate the usual content of research articles" has not been attempted, except to a small extent by the author of this article and by the ScholOnto project [11]. Hence, Section 5 listed some advantages and drawbacks of this approach – or its possible mix with the traditional approach – for researchers, reviewers and publishers. Whether or not such an approach will be adopted – or, more likely, when and under which form it will be – can only be conjectured. Furthermore, to allow a *scalable* cooperative building of a *well-organized* KB, more guidance are needed from cooperation protocols, general ontologies and domain ontologies. This seems technically achievable. As often in Semantic Web related questions, the main problem is more social than technical: will researchers use knowledge-based notations and semantic relations? If they do not, or as long as they do not, knowledge needs to be extracted automatically from informal texts, which is difficult and is also an inherently sub-optimal approach. Indeed, if information authors are not led to precise their informal terms or sentences by relating them to other ones, such information can often not be guessed by other persons, let alone software.

A related research domain is the one about the sharing of learning materials, i.e., "learning objects" (LOs). In this domain, the idea that "the smaller and less contextual the published LOs are, the better for knowledge sharing and re-use purposes" is well accepted and is mentioned in LO related standards (e.g., AICC, SCORM, and ISM). However, in this domain too, the actually published LOs are still informal documents, albeit sometimes small documents (a few paragraphs). The author of this article has proposed a fine-grained semi-formal knowledge-based approach to the sharing of learning materials and has applied it without problems to several of its own teaching courses [20].